\documentclass{rspublic}
\usepackage[dvips]{graphicx}
\usepackage{epsfig}

\begin{document}
\title[Mobility patterns in Ecology]
{Modeling the mobility of living organisms
in heterogeneous landscapes:\\
Does memory improve foraging success?}
\author[D. Boyer and P.D. Walsh]{Denis Boyer\,$^{1,2}$ and Peter D. Walsh\,$^{3}$}
\affiliation{$^{1}$Laboratoire de Physique Th\'eorique, IRSAMC, CNRS UMR 5152, Universit\'e 
de Toulouse, UPS, 31062 Toulouse, France\\
$^{2}$Instituto de F\'\i sica, Universidad Nacional Aut\'onoma de M\'exico,
D.F. 04510, Mexico\\ 
$^{3}$VaccinApe, 5301 Westbard Circle, Bethesda, MD 20816, USA}
\label{firstpage}
\maketitle

\begin{abstract}{mobility patterns, agent based models, memory,
foraging behaviour, search processes, random media}
Thanks to recent technological advances, it is now possible to track with an unprecedented 
precision and for long periods of time the movement patterns of many living organisms in their 
habitat. The increasing amount of data available on single trajectories offers the 
possibility of understanding how animals move and of testing basic movement models. 
Random walks have long represented the
main description for micro-organisms and have also been useful to understand the foraging
behaviour of large animals. Nevertheless, most vertebrates, in particular humans and other 
primates, rely on sophisticated cognitive tools such as spatial maps, episodic memory and
travel cost discounting. These properties call for other modeling approaches of mobility 
patterns. We propose a foraging framework where a learning mobile 
agent uses a combination of memory-based and random steps. We investigate how advantageous 
it is to use memory for exploiting resources in heterogeneous and changing environments.
An adequate balance of determinism and random exploration is found to maximize the foraging 
efficiency and to generate trajectories with an intricate spatio-temporal order, where 
travel routes emerge without multi-step planning.
Based on this approach, we propose some tools for analysing the non-random nature 
of mobility patterns in general.
\end{abstract}

\section{Introduction}

The literature on the movement of mobile agents has long been dominated by random 
walk models in which movement decisions are first order Markovian or with rapidly 
decaying correlations: that is, based 
entirely on conditions that are proximate in time and local in space. 
This approach has been extraordinarily successful in describing the behaviour 
of everything from microscopic particles to simple organisms such as insects and
bacteria (Turchin 1998; Colding {\it et al.} 2008). Extensions of 
random walk models beyond Brownian motion, for instance to processes governed by 
steps and/or waiting times with heavy-tailed distributions (Shlesinger $\&$ Klafter 1986), 
have also been used to describe the displacements of microzooplankton
(Bartumeus {\it et al.} 2003), spider monkeys (Ramos-Fern\'andez {\it et al.} 2004), 
marine predators (Sims {\it et al.} 2008) or even bank notes 
(Brockmann {\it et al.} 2006). Other extensions have considered multiple 
random walks,
which are useful to identify the presence of different behavioural 
states in foraging data (Morales {\it et al.} 2004)

However, the Markovian approach has serious limitations and
other research frameworks are emerging (Nathan {\it et al.} 2008; Gautestad $\&$ Mysterud 2009). 
In particular, memory storage and processing capabilities allow humans, nonhuman primates 
and other large-brained vertebrates, to transcend the shackles of a first order Markov world. 
Many animals escape the present using episodic memory, the ability not just 
to associate past events with a particular time and place but project 
how conditions at that place may have evolved since that event
(Griffiths {\it et al.} 1999; Dere {\it et al.} 2008; Rolls 2008). 
In fact, animals can even use episodic memories of
past states to predict future states (Martin-Ordas {\it et al.} 2010), a
good example being the ability of mangabeys to predict how weather has
affected the ripening of fruit (Janmaat {\it et al.} 2006$a$, $b$). 
Animals escape their current location with spatial
representations such as cognitive maps (Wills {\it et al.} 2010), Euclidean
representations of space which allow them not only to navigate
directly to important habitat features (e.g. resource patches) that
are outside of the perceptual range (Normand $\&$ Boesch 2009; Presotto $\&$
Izar 2010), but also {\it a priori} estimate the cost of traveling there
(Lanner 1996; Janson 2007; Janson $\&$ Byrne 2007; Noser $\&$ Byrne 2007).
These advanced cognitive capacities mean movements need not be
governed entirely by random decisions based on proximate states but
may also be informed by deterministic cost-benefit analysis that
compare the predicted benefits for different movement choices with
their estimated costs (Walton $\&$ Mars 2007; Hillman $\&$ Bilkey 2010).
Models motivated by observations on spider monkeys (Boyer {\it et al.}
2006), ungulates (Getz $\&$ Saltz 2008) or humans (Lee {\it et al.} 2009) have
explored how such decisions can induce complex movement patterns in
heterogeneous environments.

In the realm of Markovian processes, a lot of attention has been devoted in 
the past decade to identify efficient search strategies for finding prey or 
food patches whose locations are unknown to the forager 
(Viswanathan {\it et al.} 1999; B\'enichou {\it et al.} 2005; Viswanathan {\it et al.} 2008; 
Oshanin {\it et al.} 2009).
The impressive set of cognitive tools mentioned above offers the promise of more
efficient foraging. Nevertheless, efficiently exploiting resources still presents 
some daunting challenges. Even with perfect information about the size and location 
of resource patches, the computational load entailed in choosing the most 
efficient route through a series of patches rises exponentially with the number 
of patches, quickly becoming intractable (the classic Traveling Salesman Problem). 
What's more, in the real world information is rarely perfect. 
Resources are often ephemeral and unpredictable because of the irregularity 
of environmental forcing, resource production dynamics, and harvesting rates. 
Consequently, the accuracy of predictions about resource quality tends to 
decay with time since the last visit to a given location. Because for mobile 
agents time tends to correlate with displacement, prediction accuracy tends 
to be greatest within some limited spatial neighborhood of the agent's 
current location. This suggests that the best vertebrate foragers can hope 
for may be a combination strategy akin to that employed in search algorithms 
such as Markov Chain Monte Carlo simulations, with deterministic memory-based 
movements used to move the forager up local maxima in the rugged foraging efficiency 
landscape and occasional random steps used to explore the constantly evolving 
landscape for higher maxima.

Here we use a simulation model of monkeys foraging on fruit trees to explore 
how spatial and temporal heterogeneity in resource distribution constrain how 
advantageous it is to have a big brain. More specifically, an adequate
balance between deterministic, memory-based movement and random exploration is 
found to maximize the forager's efficiency. Therefore, some degree of stochastic 
search is still advantageous for foragers with high cognitive skills and 
foraging in not-so-scarce environments. We examine how the 
distribution of fruit tree sizes and the duration of fruiting influence
this optimal balance. We show that the concepts of efficiency and order are intricately
related in this system: optimal strategies produce
the trajectories that have the highest degree of spatio-temporal order.
Our simulation model assumes that monkeys have a set of fairly sophisticated skills, 
including cognitive maps, episodic memory, and travel cost discounting. 
Although there is mounting empirical evidence that monkeys and other vertebrates 
have these skills (Janson $\&$ Byrne 2007), our conclusions do not require them. 
Rather, they are 
also valid for a much broader class of models in which the hill-climbing 
advantages of memory are traded off against the exploratory value 
of random movement (see, {\it e.g.}, Tabone et al. 2010 in the context of 
ant foraging).

\section{Model}\label{secmodel}

{\bf Medium.} We consider a square lattice of $N\times N$ sites with lattice spacing $a$, 
where a random fraction $\rho$ of the sites is occupied by resource patches 
(fruit trees). 
A site contains 1 or 0 resource patch.
The medium is also heterogeneous in the sense that the resource patches are not 
of equal value. 
A tree located at site $i$ has a size $k_i$ representing the total amount of 
fruits it can produce each year. The tree size probability distribution function 
is given by
\begin{equation}\label{fk}
f(k)=Ck^{-\beta},\quad k_{min}\le k\le k_{max},
\end{equation}
with $k$ a continuous variable, $\beta>1$ an exponent and $(k_{min},k_{max})$ two constants. 
The form (\ref{fk}) is justified by empirical observations showing power-law distributions 
for tree sizes in tropical and temperate forests 
(Enquist $\&$ Niklas 2001; Niklas {\it et al.} 2003).  
For low values of $\beta$ (typically,
in the range $[2,4]$), a few sites concentrate an important quantity of resources, 
whereas for larger values of $\beta$ the medium is homogeneous and composed of 
trees with similar sizes.

\begin{figure}
\begin{center}
\epsfig{figure=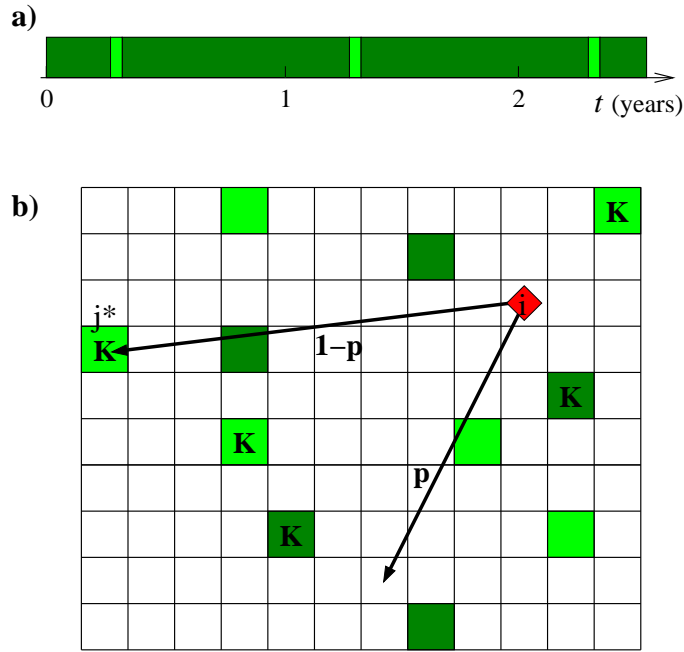,width=3.5in}
\caption{\label{Figrules} (Color online) Resource patches, {\it e.g.} fruit trees, 
are randomly distributed on a square lattice
with density $\rho$. a) Each tree produces fruit during a fruiting period (bright green) 
of duration $n_{\rm fruit}$ days and is unproductive the rest of the year (dark green). 
The date when a given tree starts fruiting is the same every year, and is randomly and 
uniformly  distributed in the interval [0, 1 year]. b) A forager located at tree $i$ 
may take a step towards a tree visited in the past (denoted with K, as in 
\lq\lq known" tree) or a random step, with probabilities $1-p$ and $p$, respectively. 
The probability $p$ involves a parameter $\eta$, see Eq.(\ref{eta}), that varies movement
between entirely random ($\eta\rightarrow\infty$) and entirely deterministic 
($\eta\rightarrow 0$).}
\end{center}
\end{figure}

Each tree produces 
fruit once a year during a fruiting period of $n_{{\rm fruit}}$ consecutive 
days. 
$n_{{\rm fruit}}(=30$ days in the following, unless otherwise indicated) 
is assumed to be the same for all trees. However, not all 
trees start fruiting  at the same time. Rather, start dates are 
randomly and uniformly distributed throughout the year, see 
Figure \ref{Figrules}a. Let us
denote $fruit_i(t)$ as the amount of ripe fruit on tree $i$ at time $t$.
During its fruiting period, $i$ produces $k_i/n_{{\rm fruit}}$ new ripe fruits 
per day ({\it i.e.}, $k_i$ fruits during the whole period).
For simplicity, the quantity  $k_i/n_{{\rm fruit}}$ becomes available 
(\lq\lq ripens") at the beginning 
of each simulation day. 
Ripe fruits remain available for a period of $n_{\rm rot}$ days after 
which they rot (become unavailable for consumption).

{\bf Forager movement rules.} A mobile agent foraging in this medium is either 
immobile eating on a fruiting lattice site with a constant feeding rate $r$ 
(eaten fruits/min) or 
moving at constant velocity $v_0$ (lattice spacings/min). The elementary 
simulation time step is $\Delta t$ ($=0.5$ min in the following).
The agent switches its activity 
from eating to moving when no ripe fruit is left in a tree, and switches from moving to
eating when the trajectory crosses a site with ripe fruits. The agent moves in a 
continuous way along linear steps joining pairs of lattice sites, and can not exit 
the $N\times N$ domain. At the end of a step or after a feeding event, the forager 
takes a new step.

Before describing the movement rules, let us note that the agent has cognitive skills 
that enable it to remember the locations, sizes and fruiting states of previously 
visited trees. Hence, during the course  of ranging, the agent adds to a list of 
known trees (marked as \lq\lq K" in Figure \ref{Figrules}b) any tree which 
is visited for the first time: the agent records the tree's location, 
size $k_i$, fruiting state (fruiting or not), and ripe fruit crop size. 
We assume that once a tree's size and location are known, they are not forgotten. 
When a known tree is revisited, the agent updates its memory about the tree's fruiting
state (for instance, from \lq\lq non-fruiting" to \lq\lq fruiting") and fruit crop size. 
We assume that each forager has a model of the fruit ripening process that 
allows it to correctly estimate how much ripe fruit should now be in a
known fruiting tree $j$, 
$fruit_j(t)$.  When a tree known by the agent to be \lq\lq fruiting" 
actually ends its fruiting period, the forager also changes the tree's state to 
\lq\lq non-fruiting" in its memory. However, foragers can only predict the 
fruiting state of trees visited during their most recent fruiting period. 
They cannot remember when the fruiting periods of trees are and, therefore, 
cannot anticipate fruiting states from one year to another.

At each movement step, the forager may or may not decide to visit a known tree.
It either heads towards a known tree $j^*$ (with probability $1-p$) or takes a random 
walk step (with probability $p$), see Figure \ref{Figrules}b. Hence, despite of its 
knowledge, the forager may decide to take a
random walk step, which is drawn from a step length 
distribution function, $f_{{\rm rnd}}(l)$, where $l>a$. 
In a deterministic step, the forager chooses as a target the known tree 
that is most efficient: {\it i.e.} that provides the best tradeoff between ripe 
fruit crop size and travel distance.
For this purpose, we introduce an one-step efficiency function, $e_i(t)$, given by:
\begin{equation}\label{ruleeff}
e_i(t)={\rm max}_{j}[F_j(t)/d_{ij}]=F_{j^*}(t)/d_{ij^*},
\end{equation}
where the index $j$ runs over the list of all known trees and $d_{ij}$ is the distance 
between forager location $i$ and tree $j$. The quantity $F_j$ is the estimated 
amount of ripe fruit in tree $j$ and, if the tree is known as \lq\lq fruiting", 
is given by
\begin{equation}\label{Fiknown}
F_j(t)=fruit_j(t),
\end{equation}
If the tree is known as \lq\lq non-fruiting" then $F_j(t)=0$ if the
tree has already been visited during the current day; otherwise,
\begin{equation}\label{Fiunknown}
F_j(t)=\langle fruit_j(t)\rangle,
\end{equation} 
the average being taken over an entire year. 
In Eq.(\ref{Fiunknown}), a tree known as \lq\lq non-fruiting" can still be 
considered as a potential destination as the tree may have started fruiting 
since the latest visit. 
Nevertheless, it tends to be less attractive than a known \lq\lq fruiting" 
tree of same size and at the same distance, as $\langle fruit_j(t)\rangle$ 
tends to be lower than the 
typical amount of fruit that $j$ carries during its fruiting period. 
The rule (\ref{ruleeff}) simply indicates that the closer and the larger 
a known tree is, the more efficient would be a step toward that tree. 
The particular tree that maximizes Eq.(\ref{ruleeff}), or the
'best' known tree, is $j^*$. Note that a fruiting tree recently depleted by the forager 
has $F_j(t)=0$ and thus will not be considered until it produces new fruits, the next day.

As explained above, the agent takes a step towards $j^*$ 
with probability $1-p$, where $p$ is given by: 
\begin{equation}\label{eta}
p=\exp(-e_i(t)/\eta),
\end{equation}
with $\eta$ a constant. The rule (\ref{eta}) indicates that the forager is 
more likely to visit $j^*$ if it is a \lq\lq good" tree (large $e_i$, low $p$). 
In turn, if the 
known trees are far away from the current location of the forager and/or of small sizes, 
the forager will prefer to take a random step instead ($p\sim1$). Therefore, the parameter $\eta$ 
does not control the memory capacities of the forager (which are assumed to be perfect), 
but the use of memory by the forager: known trees with $e_i\gg\eta$ will tend to be visited, 
while those with $e_i\ll\eta$ will be ignored. The limit
\begin{equation}
\eta \rightarrow \infty
\end{equation}
produces random walk trajectories, whereas
\begin{equation}
\eta\rightarrow0,
\end{equation}
corresponds to a forager with movements that are entirely determined by the memory, 
tending asymptotically to revisit the same fruit tree. With this set of rules, if all
known trees have been visited during a single day ({\it i.e.} in a time shorter than 
the refreshing period), then $p=1$. 
At $t=0$ the forager is placed with no previous knowledge of the medium and $p$ is 
obviously 1. Each time an unknown tree is encountered 
(during a random walk step or {\it en route} towards a known tree), it is added to the 
list of known trees, which grows as time proceeds. 

Below, we will investigate the effects of varying three parameters of the model:
$\eta$ (the tendency to take random steps), $\beta$ (the fruit crop size distribution) 
and $n_{\rm fruit}$ (the duration of each tree fruiting period). 
The eating rate ($r$), displacement speed ($v_0$) and 
$(k_{min},k_{max})$ are fixed to realistic values so that the forager 
visits a number of trees much larger than 1 on average during one day 
(the period of fruit refreshing). The forager often feeds no more than a few
minutes on a tree but may occasionally spend a few hours on a very big 
fruiting tree, which is consistent with observations on spider monkeys, for instance 
(Ramos-Fern\'andez {\it et al.} 2004).

\section{Results}

\subsection{Foraging efficiency}
\begin{figure}
\begin{center}
\epsfig{figure=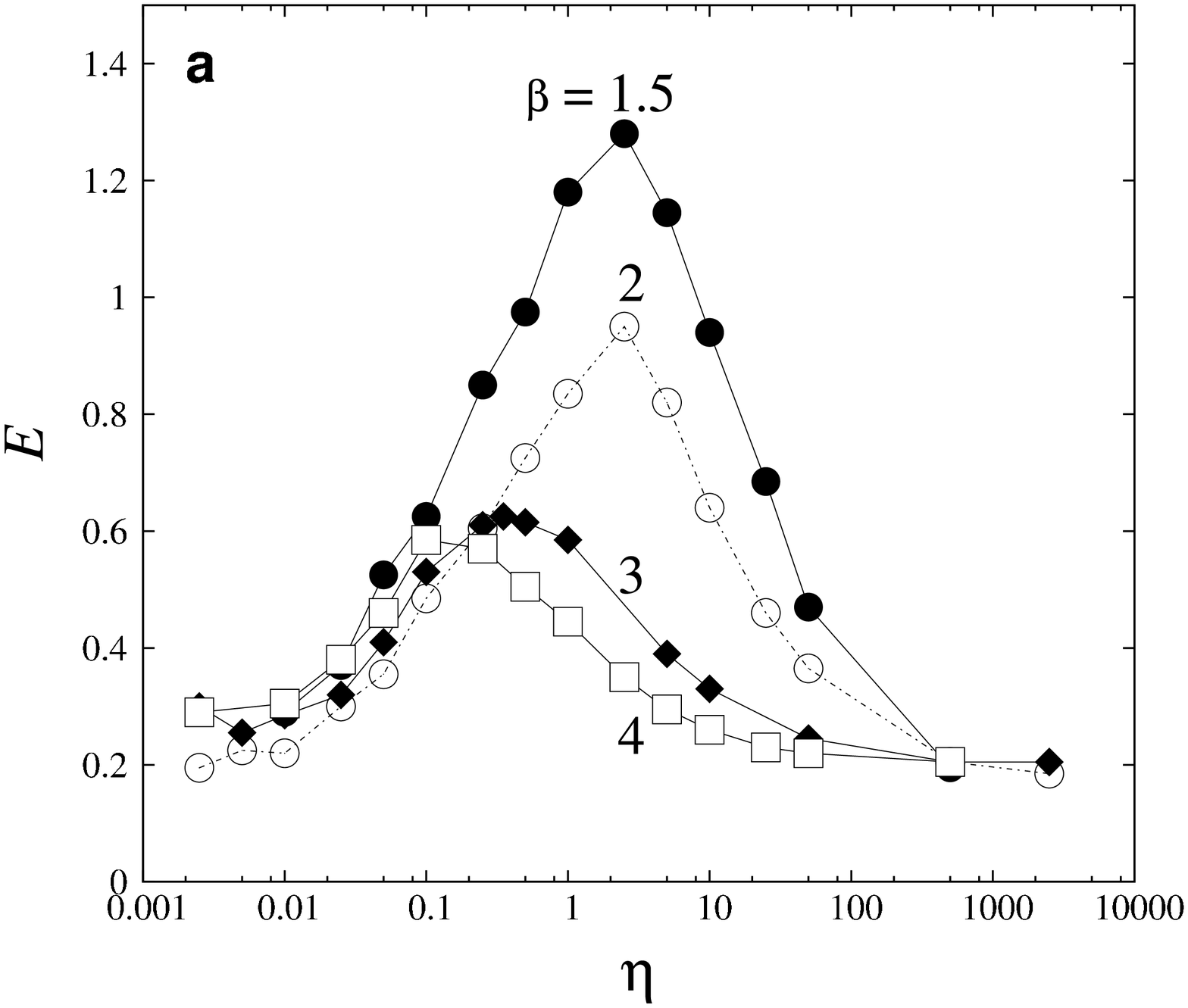,width=2.38in}
\epsfig{figure=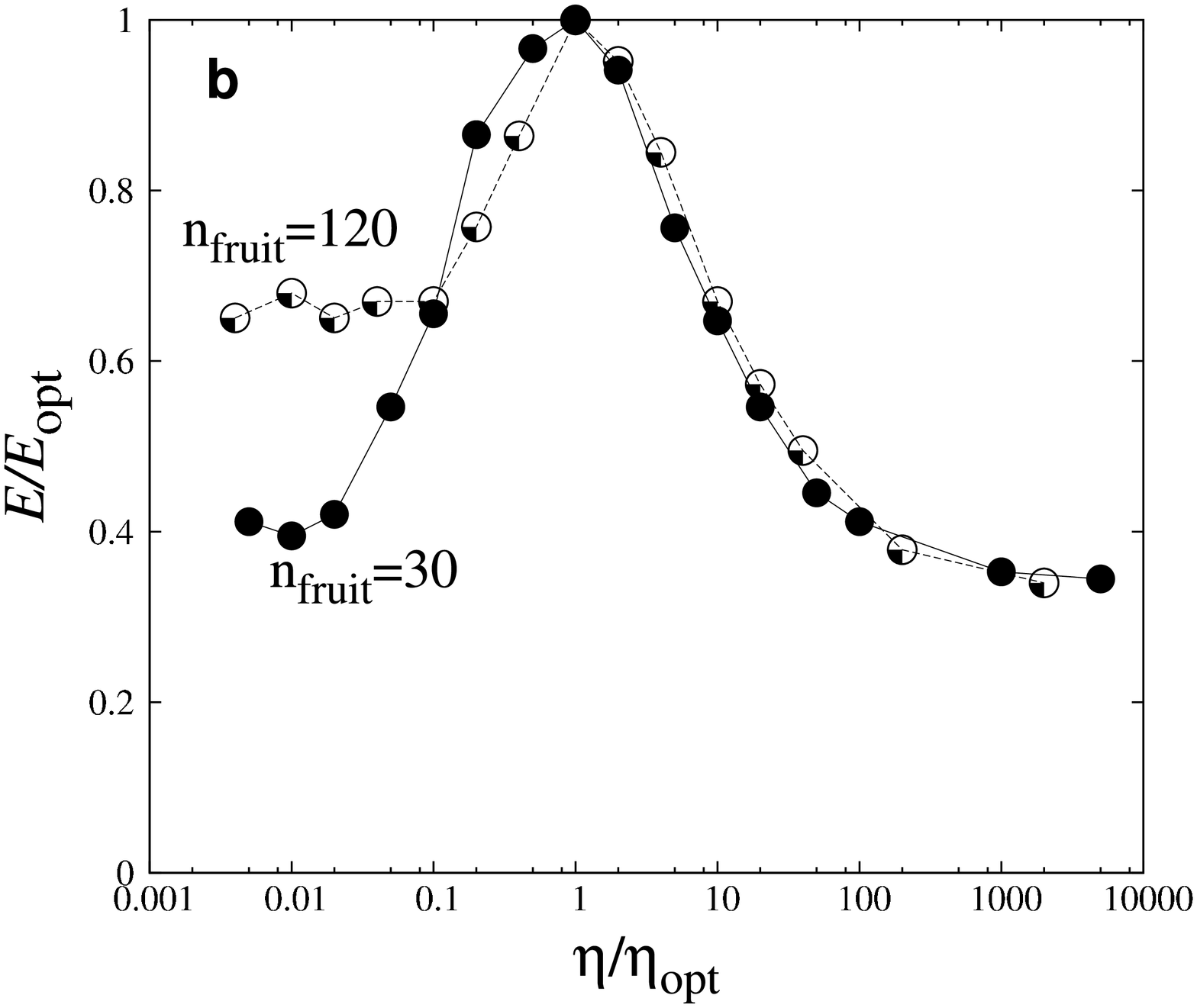,width=2.46in}
\caption{\label{Figeff} a) Foraging efficiency in media with different levels 
of fruit crop size heterogeneity ($\beta=1.5,2,3,4$). The other parameters 
are $a=1$, $\Delta t=0.5$ min., $N\times N=4\ 10^4$, $\rho=0.75$, 
$\langle k\rangle=19.9$, $k_{max}=2000$, 
$n_{fruit}=30$ days, $n_{\rm rot}=7$ days, $r=0.2$ eaten fruits/min, $v_0=2a$ min$^{-1}$, 
$f_{{\rm rnd}}(l)\propto l^{-\alpha}$ with $\alpha=2.5$ (the curves are little affected 
by other choices of $f_{\rm rnd}$). 
b) Same quantity for $\beta=3$ and two different fruiting durations, 
$n_{\rm fruit}=30$ days and $120$ days.}
\end{center}
\end{figure}

We start by computing the foraging efficiency of the mobile agent, defined as:
\begin{equation}\label{defeff}
E=\left\langle \frac{{\cal K}}{{\cal L}}\right\rangle,
\end{equation}
where ${\cal K}$ is the total amount of fruits eaten and ${\cal L}$ the total distance 
traveled by the forager. An alternate definition is $E=\langle{\cal K}\rangle/t$ 
with $t$ the foraging time, leading to similar results to those presented below. 
The average in Eq.(\ref{defeff}), which tends to a stationary value after a few years, 
is taken over different realizations of the media and foraging trajectories.
Figure \ref{Figeff}a displays $E$ as a function of the memory parameter
$\eta$, for different tree size exponents $\beta$. For easier comparison, 
$k_{min}$ is adjusted so that the total amount of resources produced in one year in
the domain, $\langle k\rangle N^2$, remains the same even though the tree size
distribution varies.

The foraging efficiency exhibits a maximum at an intermediate level of memory use, 
$\eta_{\rm opt}$. Somewhat surprisingly, the deterministic, memory-based steps 
(\ref{ruleeff}), which at small $\eta$ are increasingly common relative to random 
choices do not necessarily have a positive impact 
on the long term efficiency. This property can be qualitatively understood by noticing 
that, given the dynamical rules of Section \ref{secmodel}, random steps are often taken 
at the beginning of a run because of limited knowledge. When the number of known trees
becomes large enough, a forager with very small $\eta$ can forage exclusively on known 
trees between two fruit production events ({\it i.e.} a one day interval). 
Hence, a forager with $\eta\rightarrow0$ will tend to revisit the same set of learned 
tree locations, reducing the exploration of other areas where other fruiting trees 
could be encountered. To the contrary, a random forager ($\eta$ large) visits a 
larger number of different trees but does not use its knowledge to improve 
efficiency by moving directly to large and/or nearby trees.
Note that no fine tuning of the one-step cross-over efficiency $\eta$ that controls the 
use of random search relative to memory is necessary to bring the forager in the efficient 
region: the long term forager's efficiency varies with $\log\eta$ rather 
than with $\eta$.

A mixed strategy is the most efficient in this context, specially if resource patches have
heterogeneous sizes (small $\beta$). Whereas the foraging success of 
random and deterministic foragers is rather independent of the tree-size distribution, 
efficiency increases with medium heterogeneity in the range of intermediate strategies 
(Fig. \ref{Figeff}a). In other words, the optimal strategy becomes increasingly
advantageous with respect to others when fruits are unevenly distributed in space.
In general, the selection of the best tree $j^*$ is biased towards large tree sizes: 
{\it i.e.} thanks to rule (\ref{ruleeff}), the forager selects preferentially
a subset of trees with larger sizes than those given by a random sampling of the 
distribution (\ref{fk}). When $\beta$ decreases, this effect becomes more noticeable.
As previously shown analytically on a very simple foraging model (Boyer $\&$ Larralde 2005),
the single-step efficiency (\ref{ruleeff}) of a forager knowing many trees
tends to be higher when heterogeneity increases: on average, $F_{j^*}$ increases faster than 
the travel cost to $j^*$. This explanation does not hold, however,
at low values of $\eta$ in the present model, when the number of places susceptible 
of being visited is reduced by the lack of previous random exploration. 

Another significant aspect that favors the use of a mixed foraging strategy
is the fruiting duration $n_{\rm fruit}$ of a tree. When $n_{\rm fruit}\ll 365$, the medium 
is unpredictable to the forager: at a given time only a small fraction of the trees 
are fruiting ($n_{\rm fruit}/365=0.082$, in Fig.\ref{Figeff}a), and they fruit for 
a relatively short period of time. 
To show that behaviours exclusively based on memory ($\eta\sim 0$) are not adapted 
to rapidly changing environments, we have varied $n_{\rm fruit}$. As shown by 
Fig.\ref{Figeff}b, the use of the optimal strategy brings a greater relative advantage 
compared to more deterministic strategies when $n_{\rm fruit}$ is small.
Nevertheless, for $\eta>\eta_{\rm opt}$, {\it i.e.} for behaviours more random than
at optimality, the relative efficiency is not affected by the tree fruiting duration.

In the following sections, we investigate connections between the foraging efficiency
and the spatio-temporal structure of the walks.

\subsection{Spatial order of occupation patterns}

\begin{figure}
\begin{center}
\hspace{-0cm} \epsfig{figure=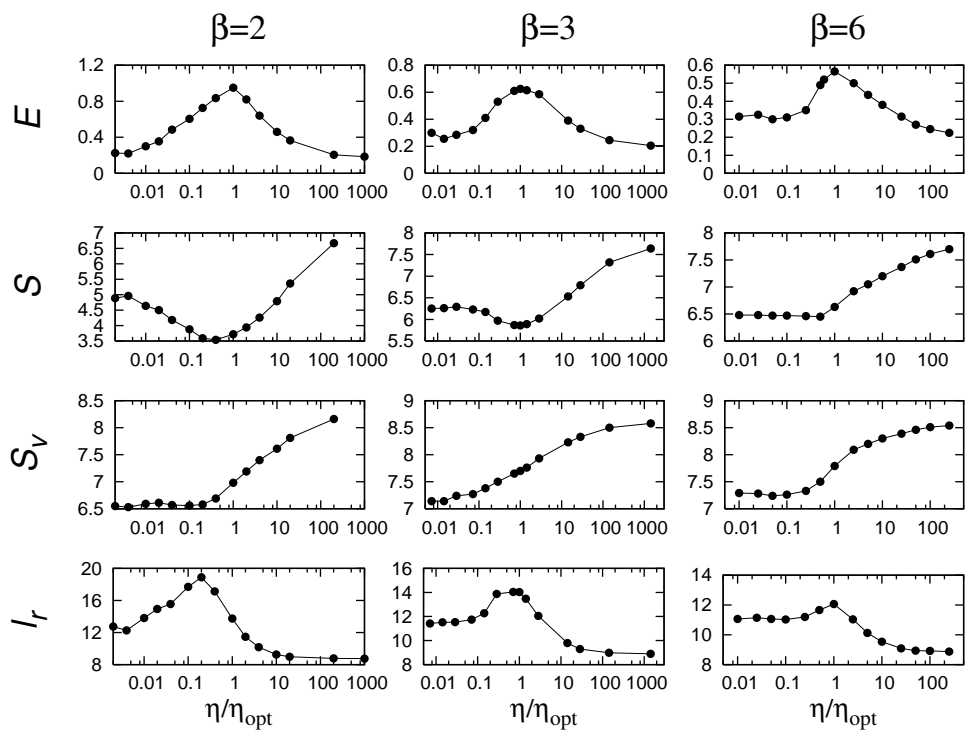,width=3.85in,angle=-90}
\vspace{0.0cm}
\caption{\label{Figentro3} Foraging efficiency ($E$), occupation entropy ($S$),
visitation entropy ($S_v$) and mean route length ($l_r$) as a function of 
the parameter
for memory use ($\eta$), for highly heterogeneous ($\beta=2$), moderately 
heterogeneous ($\beta=3$) and homogeneous ($\beta\gg1$) landscapes.}
\end{center}
\end{figure}

Processes involving memory are {\it a priori} more predictable than random walks.
Predictability has been addressed empirically in the context of human displacements 
by means 
of entropy measures (Song {\it et al.} 2010). Assume that 
the forager is optimal because it better \lq\lq knows", at any time, where the most 
profitable resource sites in the disordered environment currently are and how 
to exploit them. Do the 
corresponding walks have an ordered spatio-temporal structure that would make it easier 
for an external observer unaware of the movement rules to predict the forager position? 
Among the possible quantities related to the degree of predictability
of a trajectory, we will consider two entropies: the average occupation entropy, 
\begin{equation}
S=-\left\langle \sum_{i} p_i\ln p_i\right\rangle,
\end{equation}
where $p_i$ is the probability of finding the forager on site $i$ during a time window of
duration $T$ (the average being taken over successive time windows); and the average 
visitation entropy,
\begin{equation}
S_v=-\left\langle \sum_{i} v_i\ln v_i\right\rangle,
\end{equation}
where $v_i$ is the number of visits that site $i$ has received during a
time window of duration $T$, normalized to unity ($\sum_i v_i=1$). 
Contrary to $S$, $S_v$ does not take into account the time spent on $i$, but the number 
of steps that arrived at $i$. Figure \ref{Figentro3} display $E$, $S$ 
and $S_v$ as functions of $\eta$, for media of varying heterogeneity. 
We set $T=n_{\rm fruit}(=30$ days) as other choices give similar results.

$S_v$ has lower values when $\eta$ is low, meaning that the visits
among sites are unequally distributed. As new fruit are produced each day, the same 
fruiting trees can be revisited after being depleted. The biased choice toward larger trees
also amplify uneven visits: evident in the fact that $S_v$ decreases as $\beta$ 
decreases, {\it i.e.}, when the
probability of finding large trees in the system increases. In a given medium, 
after a plateau (in the cases $\beta=2$ and 6), the visitation entropy increases 
with $\eta$, as more frequent random steps even out the number of visits received 
by the different trees.

The behaviour of $S$ is quite different as it is not minimum at $\eta=0$ but at an 
intermediate value close to $\eta_{opt}$. In the case $\beta=3$, the location of the
minimum is actually indistinguishable from the maximum for the efficiency.
Low values of $S$ can be related to uneven 
visits and/or uneven occupation times (hence, $S<S_v$). 
The time spent at a tree is determined by its ripe fruit content.
Given the power-law distribution of the tree sizes, one may {\it a priori} expect very 
unequal feeding times in media with small $\beta$, thus lowering the entropy. 
In a given medium, the curves of $S$ show that this effect is maximal in the vicinity 
of the optimal strategy, when many of the trees visited in a given foraging time 
are currently fruiting. For $\eta<\eta_{\rm opt}$, however, the higher entropy indicates 
that residence times are more even: as confirmed by measures of ${\cal L}$ and ${\cal K}$ 
(not shown), the forager spends more time traveling and less time feeding. In this case, 
the fruiting trees are more rapidly depleted and
non-fruiting trees are more often chosen as target sites, from rule \ref{Fiunknown}. 
If a visited tree is not fruiting, the forager stays for a short time at the site, 
no matter the tree size. Similar patterns are observed at large $\eta$: in this situation, 
this is because trees are visited more randomly and most of the trees are non-fruiting.

In homogeneous media ($\beta= 6$, Figure \ref{Figentro3}), visitation and occupation
patterns become more similar. The minimum in $S$ disappears and is replaced by a flat
plateau: $S$ becomes remarkably constant for $\eta<\eta_{\rm opt}$.

\subsection{Travel routes}

The above entropy measures capture the spatial heterogeneity of the occupation
and visitation patterns but not their temporal correlations.
To gain some insight into the time regularity with which some sites are visited,
we now look for repeated sequences in the trajectories. 

As in the previous section, we decompose a whole trajectory in time windows of duration 
$T(=n_{\rm fruit}$ days in the following). We then convert the continuous forager positions 
within a same time window into a series of visited lattice sites 
$\{i_1, i_2,...,i_n\}$. As we focus here on the paths followed by the forager, 
we do not count the amount of time spent at each site (therefore, two 
consecutive sites are necessarily different). In the series of visited sites, we denote
a $k$-route as a sequence of at least $k$ consecutive sites 
which appears more than once. For practical purposes, two
sequences that are locally distant of no more than one lattice spacing $a$
are considered to be part of the same route. 
With $\rho=0.75$ and $n_{\rm fruit}=30$ 
(the values chosen for Fig. \ref{Figentro3}), 
the mean separation distance between two neighboring fruiting trees is $\sim 6a$. 
As a route is likely to join two fruiting trees, we choose $k=6$ as the minimal 
sequence length. 

\begin{figure}
\begin{center}
\epsfig{figure=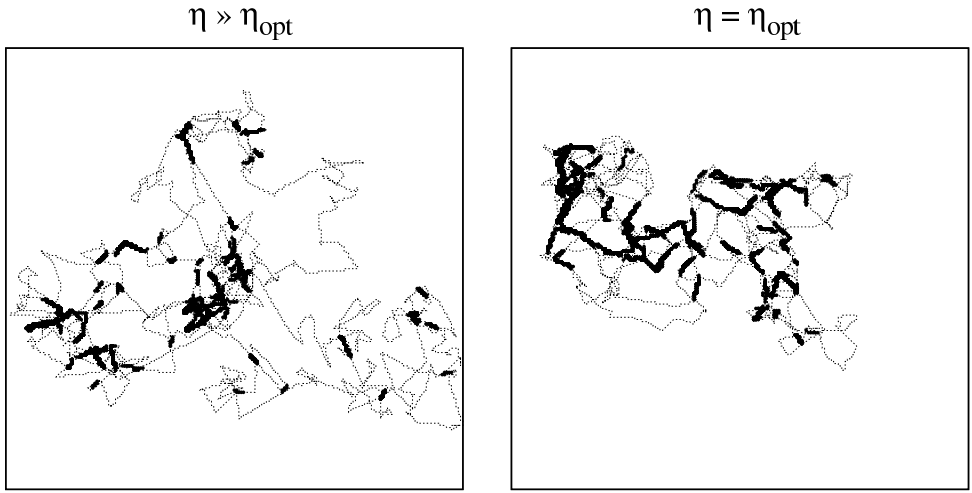,width=5in}
\caption{\label{Figroutes} One-month trajectories in a homogeneous landscape, above
and at optimality (same scale). The dark thick lines indicate the routes defined as 
subsequences of $k=6$ or more consecutive sites that appear more 
than once in the trajectory (see text for details).}
\end{center}
\end{figure}

In media with homogeneous tree sizes and for the range of values of $\eta$ considered in 
Figure \ref{Figentro3}, the 6-routes represent about 30$\%$ (higher $\eta$) to 90$\%$ 
(lower $\eta$) of a trajectory in a given time window $T$. Figure \ref{Figroutes} shows two 
trajectories with their routes highlighted. In this example, at optimality, about $52\%$ of 
the sites of a trajectory are located on the routes, the mean number 
of routes is 57 and each route is used 3.3 times on average by the forager. 
With decreasing randomness, 
the number of routes slightly decreases but each route is used much more frequently.

We denote the average length of a route as $l_r$. This quantity is analogous to a 
correlation length and indicates the persistence of the forager when it is engaged in a route. 
In other words, it measures the predictability of the forager movements when it 
starts repeating previous positions in the same order. As shown in 
Figure \ref{Figentro3},
this quantity reaches a maximum in the vicinity of $\eta_{\rm opt}$, or at the
minimum of the occupation entropy $S$. Other choices of $k$ slightly change the values obtained 
for $l_r$ but not the existence of the maximum near $\eta_{\rm opt}$. 

These results suggest that the optimal forager uses the routes as well as its persistence
along the routes to revisit several fruiting trees consecutively 
(a number of the order of $l_r/6+1\simeq 3$ to $4$). Nevertheless, the complexity of
the optimal trajectories is probably not fully described by observing a single object. 
At $\eta=\eta_{\rm opt}$, the spatial structure of the routes, which is not
captured by $l_r$ alone, reveals a relatively regular, interconnected network 
(Fig. \ref{Figroutes}). In this case, the trajectory is never very far from a route. 
By contrast, for $\eta\gg\eta_{\rm opt}$ the spatial distribution of the routes is
less uniform, some routes become isolated whereas others are clumped.

\section{Discussion and conclusion}

We have shown that trading off between random and memory-based decisions can be
advantageous for a forager searching for food. A mixed strategy is specially
rewarding if the resources are heterogeneously distributed in space and their
production period is short. Memory is in general useful, as it allows the
forager to revisit resource patches without searching. But excessive memory use
over stochastic decisions prevents the forager from updating its knowledge in rapidly 
changing environments.

Ideally, optimizing foraging efficiency in our model medium requires combined
activities: 1) to deplete nearby or large known trees, 2) to stay in the vicinity of 
known areas in order to revisit at low travel costs the depleted trees once they 
have refreshed, 3) to continuously search for new fruiting trees that will replace 
the known ones when these stop fruiting.
Finding the rules that reach the absolute maximum efficiency is a task prohibitively 
complex. Nevertheless, an animal that uses a one-step optimization procedure such 
as the one considered here (instead of many-steps planning, for instance) in 
combination with random steps can increase its efficiency by a factor of about 3 
to 7 compared with the uncombined behaviours. This efficiency gain is quite high.
In purely random search problems with memoryless agents, optimal strategies 
(obtained, for instance, by tuning the L\'evy exponent of a step length distribution) 
typically increase the efficiency by $30-40\%$ in scarce environments 
(Viswanathan {\it et al.} 1999). Notwithstanding, our results suggest that
stochastic decisions are still playing a crucial role for organisms 
with extremely high cognitive skills and foraging in not-so-scarce environments. 
These findings open other perspectives in the debate on the relevance of random searches 
in Biology (Bartumeus $\&$ Catalan 2009; Smouse {\it et al.} 2010).

Memory leads to unexpected emergent patterns of habitat use, many of which
still need to be studied. As shown by a recent multiple random walk approach,
revisits are not independent events when memory effects are at play (Gautestad $\&$ Mysterud 2009).  
In our model, high efficiencies coincide with movements
of higher predictability, or order. We find that order is enhanced in environments 
with broad tree size distributions, where a few patches concentrate many resources.
Low values are found for an entropy based on
time-averaged spatial occupation patterns. Trajectories
also become more predictable from a dynamical point of view,
with longer repeated sequences of visited sites.
Similar measures have revealed the regular nature of human displacements 
(Song {\it et al.} 2010) and their application to real ecological data could be useful
to unveil how animals exploit resources. In this context,
predictability also has implications on the risk of predation: if one associates low 
forager's entropy levels with higher predation risks, a deviation from optimal
foraging may decreases that risk. This may be achieved by increasing the randomness 
of the displacements, for instance.

At optimality, the model forager spends about half of its traveling time 
revisiting previous places in an orderly way, an activity which is reminiscent 
of the travel routes used by real animals (Noser $\&$ Byrne 2007). 
In fact, observation that monkeys
persistently visit the same sequence of trees has been used to
argue that monkeys navigate by choosing from a limited set of
multi-step routes rather than Euclidean cognitive map calculations 
(Di Fiore $\&$ Suarez 2007).  What our simulations show is that
persistent use of the same sequence of trees need not imply the use of
a pre-ordained, multi-step route. Rather, what looks like a multi-step
route can emerge spontaneously from a one step optimization algorithm.
In areas that are well-known, the forager's cognitive map of
the size and location of trees is essentially the same at every visit.
As the forager tends to make the same sequence of decisions each
time it visits an area, its episodic memory of local
fruiting states is also similar between visits. This might occur if
depletion dynamics resulted in a characteristic return
time such that trees tended to be at the same state of refresh at each
visit. Or characteristic return time might be so long that fruiting
states were poorly known. In both cases the predicted value for each
tree would be proportional to its size and the relative values of
alternative choices at each step of the chain would be consistent
between visits.

Directional persistence is another observation often
interpreted to support multi-step route planning over the use of one step, Euclidean
cognitive map navigation. Here,
model foragers also tended to show a high level of directional
persistence, continuing on in the same direction after feeding on a
tree. Directional persistence occurred in our
simulations because a forager {\it en route} to a high yield large tree often
passed close enough to a smaller tree to make a short detour cost
effective (Janson 2007).

These examples raise a much broader issue in the study of animal
movement. Many previous empirical studies have attempted to test
alternative models of animal navigation in terms of relatively simple,
intuitive predictions such as \lq\lq follows route or not" or \lq\lq takes direct
path or not". What the example above illustrates is not just that
alternative models often make the same prediction (Janson $\&$ Byrne
2007), but that the fact that a given alternative model predicts a
particular movement pattern is not always intuitively obvious. That a
one step optimization algorithm can cause repeated or directionally
persistent sequences of movement is not a foregone conclusion that
flows necessarily from the algorithm's assumption. It is an emergent
consequence of the algorithm's application in a heterogeneous
environment.

 More importantly, just as alternative models can non-intuitively
predict the same emergent patterns, alternative models can also
predict very different emergent patterns that often cannot be
anticipated from the decision algorithms themselves. These emergent
patterns represent a huge and virtually untapped body of data for
discriminating between animal navigation mechanisms and motivations.
However, what emergent patterns that are predicted by complex
navigation models such as the one we describe can only be evaluated
through simulation. This necessitates a shift in statistical inference
towards methods such as Approximate Bayesian Computation (ABC), which
allows the fitting of arbitrarily complex simulation models to real
field data (Beaumont {\it et al.} 2002). What makes ABC particularly
attractive for inference on movement models is the fact that it can
use observations not just on the turning angle or velocity of
particular path segments, but on multiple ensemble properties of
movement such as the mean, variance or autocorrelation of angles or
velocities and their cross correlation with environmental features. We
expect ABC and similar methods to play pivotal roles in shifting the
axis of research on animal movement from theory and experiment towards
field observation.

\begin{acknowledgements}

This work was conducted as a part of the \lq\lq Efficient Wildlife Disease Control: 
From Social Network Self-organization to Optimal Vaccination" 
Working Group supported by the National Center for Ecological Analysis and Synthesis, 
a Center funded by NSF (Grant $\#$DEB-0553768), the University of California, 
Santa Barbara, and the State of California. We thank the group members,
J. Benavides, D. Caillaud, M. C. Crofoot, W. M. Getz, K. Hampton, L. A. Meyers, 
S. J. Ryan, L. Salvador and S. V. Scarpino for discussions and collaborations 
on this subject.

\end{acknowledgements}

\label{lastpage}
\end{document}